\documentclass[12pt,a4paper]{article}
\usepackage{amsmath,amssymb,graphicx,indentfirst}
\usepackage{amssymb}
\usepackage{hyperref}
\numberwithin{equation}{section}
\allowdisplaybreaks[3]

\begin{document}
%%% Title page %%%%%
\begin{titlepage}

 \renewcommand{\thefootnote}{\fnsymbol{footnote}}
\begin{flushright}
 \begin{tabular}{l}
 RUP-16-1\\
% arXiv:1601.01784\\ %This should be replaced after submission.
%\today %This should be commented out.
 \end{tabular}
\end{flushright}

 \vfill
 \begin{center}

% \vskip 2.5 truecm

\noindent{\Large \textbf{The masses of higher spin fields on AdS$_4$}}

\medskip

\noindent{\Large \textbf{and conformal perturbation theory}}
\vspace{1.5cm}

\noindent{Yasuaki Hikida}
\bigskip

 \vskip .6 truecm

\centerline{\it Department of Physics, Rikkyo University, Toshima, Tokyo 171-8501, Japan}

\bigskip

\centerline{\tt hikida@rikkyo.ac.jp}

 \end{center}

 \vfill
\vskip 0.5 truecm

\begin{abstract}

We study the breaking of gauge symmetry for higher spin theory on AdS$_4$  dual to the 3d critical O$(N)$ vector model. 
It was argued that the breaking is due to  the change of boundary condition for a scalar field through a loop effect and the Goldstone modes are bound states of a scalar field and higher spin field. 
The masses of higher spin fields were obtained from the anomalous dimensions of dual currents at the leading order in $1/N$,
and we reproduce them from the O$(N)$ vector model in the conformal perturbation theory.
The anomalous dimensions can be computed from the bulk theory using Witten diagrams, and we show that the bulk computation reduces to the boundary one in the conformal perturbation theory.
With this fact our computation provides an additional support for the bulk interpretation.

\end{abstract}
\vfill
\vskip 0.5 truecm

\setcounter{footnote}{0}
\renewcommand{\thefootnote}{\arabic{footnote}}
\end{titlepage}

\newpage

\tableofcontents
%%%%%%%%%%%%%%%%%%%%%%%%%%%%%%%%%%%%%%%%%%%%%%%%%%%%%%%%%%%%%%%%%%%%%%
%\newpage

\section{Introduction}

Superstring theory includes a large amount of massive higher spin states, and higher spin gauge symmetry is expected to appear at the tensionless limit.
This implies that superstring theory with finite tension could be described by higher spin gauge theory with its symmetry broken \cite{Gross:1988ue}.
Recently, a large progress has been made by working on the AdS space, where we can utilize the Vasiliev theory \cite{Vasiliev:2003ev} and the AdS/CFT correspondence.
The first concrete example of AdS/CFT with Vasiliev theory was proposed by
Klebanov and Polyakov \cite{Klebanov:2002ja} (see also \cite{Sezgin:2002rt}), where the
4d minimal bosonic Vasiliev theory \cite{Vasiliev:1995dn,Vasiliev:1999ba}
is dual to the 3d O$(N)$ vector model.
In this paper, we study the breaking of higher spin gauge symmetry of the Vasiliev theory as the most basic example.
We apply the method developed in \cite{Creutzig:2015hta} for lower dimensional dualities, where related works may be found in \cite{Hikida:2015nfa,Gaberdiel:2015uca,Gwak:2015jdo}.

Our motivation to study the symmetry breaking is to understand  the relation between superstring theory and higher spin gauge theory.
Extending the duality of \cite{Klebanov:2002ja}, 
the authors of \cite{Chang:2012kt} proposed a concrete relation via 3d ABJ(M) theory in \cite{Aharony:2008ug,Aharony:2008gk}, where
the relation is named as ABJ triality.
A lower dimensional analogue of \cite{Klebanov:2002ja} was conjectured in \cite{Gaberdiel:2010pz}.
Based on the duality, lower dimensional versions of the ABJ triality  were proposed in \cite{Gaberdiel:2013vva,Gaberdiel:2014cha,Gaberdiel:2015mra,Gaberdiel:2015wpo} with large or small $\mathcal{N}=4$ supersymmetry and in  \cite{Creutzig:2014ula,Hikida:2015nfa} (see also \cite{Creutzig:2011fe,Creutzig:2013tja}) with $\mathcal{N}=3$ supersymmetry.
In \cite{Hikida:2015nfa,Creutzig:2015hta}, the breaking of higher spin symmetry has been studied in the $\mathcal{N}=3$ holography, but it seems that a deeper understanding is necessary to say something concrete about the relation to superstring theory.
For this purpose it should be useful to examine the most basic example of \cite{Klebanov:2002ja}.

In \cite{Klebanov:2002ja}, they considered a 4d  Vasiliev theory with a scalar field along with higher spin gauge fields with even spin, and we can assign the Dirichlet or Neumann boundary condition to the scalar field.
The 4d Vasiliev theory with the Neumann boundary condition is proposed to be dual to the free 3d O$(N)$ vector model with the O$(N)$ invariant condition.
The operator $\mathcal{O}$ dual to the scalar field has the scaling dimension $\Delta = 1$.
We can consider the RG flow included by the following double trace deformation as
\begin{align}
 \Delta S = \frac{f}{2} \int d^3 x \, \mathcal{O} (x) \mathcal{O} (x) \, .
 \label{def}
\end{align}
The deformation is argued to be dual to the change of boundary condition of bulk scalar field \cite{Witten:2001ua}.  In particular, the critical theory at the IR fixed point should be dual to the Vasiliev theory with the Dirichlet (or $\Delta = 2$) boundary condition.

The higher spin symmetry is broken at the order of $1/N$ at the critical point.
Therefore, we can expect that the bulk higher spin symmetry is broken due to the change of boundary condition,
and the Higgs mass is generated through a one-loop effect. 
In fact, it was shown in \cite{Girardello:2002pp} by group theoretical analysis that 
 Goldstone modes are bound states of scalar field and higher spin field.
The fields on Euclidean AdS$_4$ can be classified with its isometry SO$(4,1)$ or its dual conformal symmetry.
We use quantum numbers under its bosonic subalgebras as $(\Delta ,s )$, where $\Delta$ and $s$ are the dual scaling dimension and the spin, respectively. The unitarity boundary is given by $\Delta \geq s + 1$,
and the representation $D (\Delta . s)$ decouples at the limit $\Delta = s+1$ as
\begin{align}
\lim_{\Delta \to s + 1} D (\Delta , s) \to 
D(s+1,s) \oplus D(s+2,s-1) \, .
\end{align}
This is related to the fact that dual CFT current with spin $s$ becomes conserved as $\partial \cdot J_s = 0$ at the limit.
In order words, the short representation $D(s,s+1)$ becomes long by eating a long representation 
$D(s+2,s-1)$.
The Vasiliev theory only includes higher spin gauge fields and a scalar field, so we do not have such fields.
However the representation $D(s+2,s-1)$ can be realized as bound states
of spin $s'$ gauge field $(s > s' \geq 2)$ and the scalar field with $\Delta = 2$.
This can be seen from 
\begin{align}
D(s ' +1  , s ' ) \otimes D(2,0) = \bigoplus_{S=0}^\infty \bigoplus _{n=0}^\infty D (s ' + S + n + 3 , s ' + S )  \, ,
\label{bound}
\end{align}
where  $D(s+2,s-1)$ is realized for  $(S,n) = (s - s' - 1 , 0)$.
Therefore we can conclude that the Goldstone modes are bound states of gauge field of spin $s'$ $(< s)$ and the scalar field with $\Delta = 2$.
Notice that there is no such $(S,n)$ for spin $s=2$ field, so the graviton is kept massless.

The Higgs mass $M_{s}$ of spin $s$ field is easier to compute from the dual critical O$(N)$ vector model. 
The mass can be obtained from the scaling dimension $\Delta_{s}$ of dual current $J_s$ by using the map
\begin{align}
M^2_{s} = \Delta_{s} (\Delta_{s} - 3) - (s - 2) ( s + 1) \, .
\label{dictionary}
\end{align}
The anomalous dimension $\tau_s \equiv \Delta_{s} - s - 1$  after the deformation \eqref{def} was obtained purely within the critical O$(N)$ vector model as \cite{Ruhl:2004cf}% 
\footnote{While completing this paper, we become aware of  \cite{Skvortsov:2015pea,Giombi:2016hkj}, where anomalous dimensions (or those for $d$-dimensions in \cite{Lang:1990re,Lang:1992zw}) were  reproduced from the critical model utilizing the method developed in \cite{Rychkov:2015naa}. The method is different from the one in \cite{Ruhl:2004cf} and ours.
}
\begin{align}
 \tau_s = \frac{16 (s-2)}{3 \pi^2 N (2s - 1)} 
 \label{Ruhlr}
\end{align}
at the leading order in $1/N$. The dictionary \eqref{dictionary} thus leads to
\begin{align}
 M^2 _{s} = \frac{16 (s - 2) }{3 \pi^2 N } \, .
 \label{masss}
\end{align}
The aim of this paper is to develop a method to rederive \eqref{masss} in a way such that we can improve our understanding of the symmetry breaking from the bulk viewpoint.

\begin{figure}
  \centering
  \includegraphics[width=4.5cm]{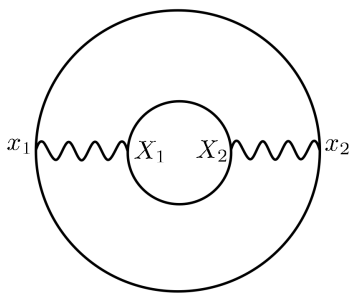}
  \caption{The Witten diagram corresponding to the one-loop contribution of current-current two point function.}
  \label{2pt}
\end{figure}

Now we have the following bulk picture of the symmetry breaking; 
\begin{itemize}
\item[(1)] The symmetry breaking is due to the change of boundary condition for scalar field through a loop effect.
\item[(2)] The Goldstone modes are bound states of higher spin gauge field and the scalar field.
\end{itemize}
 A direct way to confirm them is to read off the Higgs masses from one-loop corrections to higher spin propagators as is done for massive graviton, e.g., in \cite{Porrati:2001db,Duff:2004wh}. 
However, the spin 2 computation was quite complicated, and it looks difficult to generalize the analysis to higher spin gauge fields, see \cite{Manvelyan:2008ks} for a previous work.
Instead of reading the masses from bulk-to-bulk propagators, we compute the anomalous dimensions of dual current from boundary two point functions via the bulk Witten diagrams with the boundary-to-boundary propagators.
With this form it becomes easier to read off quantum numbers corresponding to the masses utilizing dual conformal symmetry (or AdS isometry).
The bulk interpretation (1) suggests that the anomalous dimensions can arise from the diagram in fig. \ref{2pt}
with the scalar field with Dirichlet boundary condition running along a line of the loop.

\begin{figure}
  \centering
  \includegraphics[width=10cm]{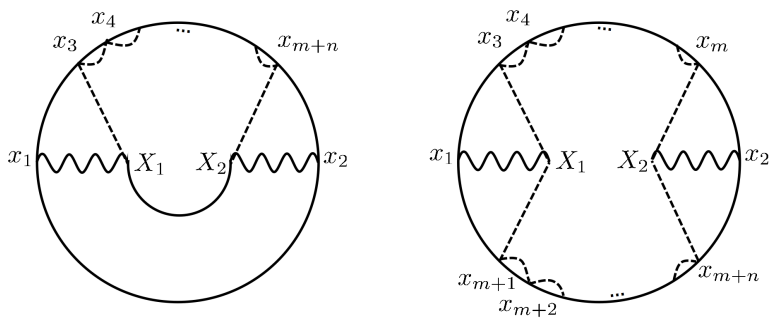}
  \caption{The Witten diagrams with $(m+n-2)$ insertions of boundary deformation operator.}
  \label{pert}
\end{figure}

The loop in the diagram is still difficult to evaluate, so we use another trick to simplify the computation.
As mentioned above, the change of boundary condition corresponds to the insertions of boundary deformation operator \cite{Witten:2001ua}.
We can show that the shift in the bulk-to-bulk propagator can be reproduced by summing all possible insertions and integrating all positions of inserted operator.
Thus, the contributions to anomalous dimensions come from the diagrams with boundary insertions as in fig. \ref{pert}.
These diagrams consist of bulk Witten diagrams at the tree level, and they are known to be reproduced by free boson theory.%
\footnote{Correlation functions in the free boson theory can be evaluated by applying the Wick contraction (see, e.g., \eqref{npt}). 
Three point functions were reproduced from the bulk theory in  \cite{Giombi:2009wh,Giombi:2010vg}. Moreover, it was shown in \cite{Maldacena:2011jn} that the correlation functions in 3d conformal field theory with higher spin symmetry are the same as those in the theory of  free bosons or fermions.	Generic $N$-point functions were obtained from the bulk Vasiliev theory in \cite{Didenko:2012tv}, see also \cite{Gelfond:2013xt,Didenko:2013bj}.
}
In this way, we map the computation of bulk Witten diagram with one-loop correction to the boundary one. In fact, the computation can be done by the O$(N)$ vector model in conformal perturbation theory with large $N$ factorization.

In this paper, we reproduce the anomalous dimensions of dual currents in terms of conformal perturbation theory. As explained above, this should confirm the bulk interpretation (1). 
The other interpretation (2) suggests that the anomalous dimension of spin $s$ current arises only from the diagrams where the gauge field of spin $s' (< s)$ runs along with a line in the loop and a scalar runs along the other line. Since we do not evaluate directly the loop diagram in fig.~\ref{2pt},
we cannot rule out the possibility of other contributions to the anomalous dimension.
Therefore, in this sense, our computation does not completely confirm the other interpretation (2).

The rest of this paper is organized as follows;
In the next section, we explain the basic facts on the free O$(N)$ vector model and conformal perturbation theory. We then show that computations in the conformal field theory can be interpreted from the bulk theory by identifying integrals in boundary computations to bulk Witten diagrams. Furthermore, we identify  integrals we need to compute and summarize our results.
In section \ref{computation}, we compute the integrals explicitly by using a way of regularization.
We conclude this paper and discuss future problems in section \ref{conclusion}.
In appendix \ref{AppA}, we summarize the integral and sum formulas used during the computations.
In appendix \ref{AppB},  we show that the computation from the bulk Witten diagrams reduces to that in the boundary conformal perturbation theory.

\section{Methods}
\label{methods}

In this section, we explain how to compute the anomalous dimensions from the boundary theory in conformal perturbation theory. At the same time, we give
the bulk interpretation of the computation in terms of Witten diagrams more explicitly.

\subsection{Preparations}

We consider the theory of $N$ free bosons $\phi_i$ $(i = 1,2,\ldots , N )$ and  deform the theory as \eqref{def} with $\mathcal{O}=\phi_i \phi^i$.
The critical theory is obtained by taking the limit of $f \to \infty$.
In the conformal perturbation theory, correlation functions after the deformation can be computed as
\begin{align}
\left \langle \prod_{i=1}^n \Phi_i (x_i) \right \rangle _f
 = \frac{\left \langle\prod_{i=1}^n \Phi_i (x_i)  e^{- \Delta S} \right \rangle _0 }{\left \langle e^{- \Delta S} \right \rangle _0} \, .
 \label{defpert}
\end{align}
Here $\Phi_i$ are some operators and the correlators with subscript $0$ are computed in the free theory. In this way, we can compute the correlation functions after the deformation in terms of those in the free theory.

The free O$(N)$ vector model has conserved currents 
$J_{\mu_1 \cdots \mu_s} (x)$ with even $s$, where the indices are symmetric and traceless. 
Introducing polarization vector $\epsilon$, we define 
$J_s(x ; \epsilon) \equiv J_{\mu_1 \cdots \mu_s} (x) \epsilon^{\mu_1} \cdots \epsilon^{\mu_s} $.
Using the traceless condition, we can set $\epsilon \cdot \epsilon = 0$.
As in \cite{Giombi:2009wh}, we define generating function as
\begin{align}
\mathcal{O} ( x ; \epsilon)
 = \sum_{s=0}^\infty J_s (x ; \epsilon)
  = \phi_i ( x) f(\epsilon \cdot \overrightarrow{\partial} , \epsilon \cdot \overleftarrow{\partial}) 
  \phi^i ( x) \, ,  \quad  f ( u , v) = e^{ u -  v} \cdot
    \cos \left(2 \sqrt{u  v  }\right) \, .
\end{align}
Using the Wick contraction of free scalar fields, we can compute the $n$-point correlator of generating function as
\begin{align}
\label{npt}
& \left  \langle \prod_{i=1}^n \mathcal{O} (x_i ;  \epsilon_i) \right \rangle_0 = \frac{2^{n-1} N}{n}  \\
& \qquad  \qquad \times  \sum_{\sigma \in S_n} P_\sigma
  \prod_{i=1}^n \left[  \cos \left(2\sqrt{\epsilon_i  \cdot  \overleftarrow{\partial}_i  \epsilon_i \cdot  \overrightarrow{\partial}_i } \right)  \frac{1}{|x_i - x_{i+1} + \epsilon_i + \epsilon_{i+1}|} \right] \, .
  \nonumber 
 \end{align}
Here $P_\sigma$ denote the permutation of $(x_i;\epsilon_i)$ by $\sigma \in S_n$.

Using this expression,
the two point function of higher spin current $J_s$ with $s \geq 2$ can be computed as (see (4.102) of \cite{Giombi:2009wh})
\begin{align}
\langle J_s (x_1 ; \epsilon_1) J_s (x_2 ; \epsilon_2 ) \rangle 
_0  =  N_s \frac{ (x_{12}^-)^{2s} }{|x_{12}|^{4s+2}}  \, , \quad
N_s = \frac{ N  (2s)!}{(s!)^2} \, , 
\label{before2pt}
\end{align}
where we have set $\epsilon_1 = \epsilon_2$ and used $x_{12}^- 
= 2 \epsilon_1 \cdot (x_1 - x_2) = 2 \epsilon_1 \cdot x_{12}$.
The two point function of the scalar operator $\mathcal{O} \equiv J_0$ is
\begin{align}
\langle \mathcal{O}(x_1) \mathcal{O}(x_2) \rangle _0 = \frac{2N}{|x_{12}|^2} \, , \quad
\langle \mathcal{O}(k_1) \mathcal{O}(k_2) \rangle _0 = G(k_1) \delta^{(3)}(k_1 + k_2) \, , 
\quad 
G(k) = \frac{4 \pi ^2  N  }{|k|}  \, .
\label{scalar2pt}
\end{align}
It will be  useful to move to the momentum basis 
\begin{align}
 \mathcal{O}(k) = \frac{1}{(2 \pi)^{3/2}} \int d^3 x \, \mathcal{O}(x) e^{i k \cdot x}
 \label{momentum}
\end{align}
by using the formulas \eqref{Fourier} and \eqref{formula}.

We can easily compute the two point function of the scalar operator after the deformation as
\begin{align}
\langle \mathcal{O}(k) \mathcal{O}(-k) \rangle _f
 =  G(k) - f  G(k)^2 + f^2 G(k)^3  + \cdots 
 = \frac{G(k)}{1 + f G(k)} 
\end{align}
using 
\begin{align}
 \Delta S = \frac{f}{2} \int d^3 k \, \mathcal{O} (k) \mathcal{O} (-k) \, .
 \label{defm}
\end{align}
 This two point function can be reproduced from the bulk theory with scalar field on AdS$_4$ \cite{Witten:2001ua}, see also \cite{Mueck:2002gm}.
We are mainly interested in the IR limit with $f \sim \infty$, where we have
\begin{align}
\langle \mathcal{O}(k) \mathcal{O}(-k) \rangle _f
 \sim \frac{1}{f} - \frac{1}{f^2} G(k)^{-1} 
\end{align}
or
\begin{align}
\langle \mathcal{O}(x_1) \mathcal{O}(x_2) \rangle _f
\sim  \frac{1}{f} \delta ^{(3)} (x_{12}) + \frac{1}{f^2} \frac{1}{4 \pi^4 N} \frac{1}{|x_{12}|^4} 
\label{sum1}
\end{align}
 with the coordinate basis.
 Neglecting the contact term, we reproduce the two point function of scalar operator with
 $\Delta = 2$.
We will also need 
\begin{align}
 G(x_{12})_f \equiv
\delta ^{(3)} (x_{12}) - f \langle \mathcal{O}(x_1) \mathcal{O}(x_2) \rangle _f 
\sim  - \frac{1}{f} \frac{1}{4 \pi^4 N} \frac{1}{|x_{12}|^4} 
\label{sum2}
\end{align}
in the following analysis.

\subsection{Current-current two point functions}

In the conformal perturbation theory, the two point function of higher spin current with generic $f$ can be computed in the free theory as
\begin{align}
\nonumber 
  & \langle J_s (x_1 ; \epsilon_1)  J_s  ( x_2 ;  \epsilon_2)  \rangle_f
   =   \langle  J_s (x_1 ; \epsilon_1) J_s  ( x_2 ;  \epsilon_2)  \rangle_0 \\
   & \qquad \qquad -\frac{f}{2}\int d^3 x_3 \langle  J_s ( x_1 ; \epsilon_1) J_s (x_2 ;  \epsilon_2)   \mathcal{O} ( x_3) \mathcal{O}( x_3)\rangle_0 
    \\ & \qquad \qquad + \frac{f^2}{8} \int d^3 x_3 d^3 x_4 \langle  J_s  (x_1 ; \epsilon_1) J_s (x_2 ; \epsilon_2)   \mathcal{O} ( x_3) \mathcal{O} ( x_3) \mathcal{O} (x_4) \mathcal{O} (x_4)\rangle_0
+ \cdots \nonumber 
\end{align}
with the contributions from the denominator of \eqref{defpert} extracted.
At the free limit with $f=0$, the two point function of higher spin current is given as \eqref{before2pt}.
At the IR fixed point with $f \to \infty$, there will be contributions $\delta_s , \tau_s$ at  the order  $N^{-1}$ as
\begin{align}
\label{anomalous}
\langle J_s ( x_1 ; \epsilon_1) J_s ( x_2 ;  \epsilon_2 ) \rangle 
_{f \to \infty}  &= N_s ( 1 + \delta_s) \frac{ (x_{12}^-)^{2s} }{|x_{12}|^{4s+2 + 2 \tau_s}}  + \mathcal{O}(N^{-1})  \\
 &=N_s  \frac{ (x_{12}^-)^{2s}}{|x_{12}|^{4s+2 }} (1 + \delta_s - 2 \tau_s \log |x_{12}|)  
 + \mathcal{O}(N^{-1})   \nonumber 
\end{align}
with $N_s \propto N$.
Here $\tau_s$ is the anomalous dimension, while $N_s \delta_s$ is the change of normalization.
We are interested in the anomalous dimension, so we will concentrate on the contribution proportional to $\log |x_{12}|$.

Since we know that there is no contribution to the anomalous dimension from the zeroth order term in $f$, we start to examine the first order term.
The four point function in the integral is written as
\begin{align}
& \langle  J_s ( x_1 ; \epsilon_1 ) J_s ( x_2 ; \epsilon_2) \mathcal{O} ( x_3) \mathcal{O} ( x_3) \rangle _0  = 
16 N  \frac{1}{|x_{31}|} 
e^{ - \epsilon_1 \cdot \overleftarrow{\partial}_1 } 
 \cos \left( 2 \sqrt{ \epsilon_1 \cdot  \overleftarrow{\partial}_1 \epsilon_1 \cdot  \overrightarrow{\partial}_1}\right) 
e^{  \epsilon_1 \cdot \overrightarrow{\partial}_1 } 
 \frac{1}{|x_{13}|} \nonumber  \\ &  \qquad \qquad \times  \frac{1}{|x_{32}|}
 e^{ - \epsilon_2 \cdot \overleftarrow{\partial}_2 } 
 \left.  \cos \left( 2 \sqrt{ \epsilon_2 \cdot  \overleftarrow{\partial}_2 \epsilon_2 \cdot  \overrightarrow{\partial}_2}\right) 
 e^{  \epsilon_2 \cdot \overrightarrow{\partial}_2 } 
 \frac{1}{|x_{23}|} \right|_{\epsilon_1^s \epsilon_2^s} \, ,
\end{align}
which can be obtained from \eqref{npt}.
Thus the first order term can be given by derivatives of the following integral
\begin{align}
\int d^3 x_3 \frac{1}{|x_{13}|^2 |x_{23}|^2} =  \frac{\pi^3}{|x_{12}|}
\end{align}
with respect to $x_1, x_2$. The integral has been computed by applying the formula \eqref{line}.
Since there is no term proportional to $\log |x_{12}|$,
we can conclude that there is no contribution to the anomalous dimension from the first order term.

\begin{figure}
  \centering
  \includegraphics[width=10cm]{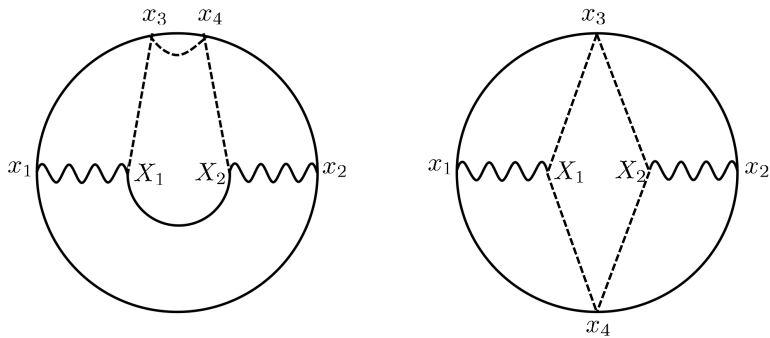}
  \caption{The Witten diagrams corresponding to the $f^2$ order contributions to the current-current two point function.}
  \label{critical0}
\end{figure}

Next we move to contributions of higher order in $f$ but still at the leading order in $1/N$.
From the order of $f^2$, there are two types of contributions as%
\footnote{The factor $2 \cdot 2$ comes from the choice of $\mathcal{O}$ in \eqref{def}.}
\begin{align}
\label{type1}
\tilde I_1 = \frac{f^2}{2}
\int d^3 x_3 d^3 x_4 \langle  J_s  ( x_1 ;  \epsilon_1)  J_s ( x_2 ;  \epsilon_2)   \mathcal{O} ( x_3) \mathcal{O} ( x_4)  \rangle _0 \langle  \mathcal{O}( x_3) \mathcal{O}( x_4)\rangle_0
\end{align}
and
\begin{align}
\label{type2}
\tilde I_2 = \frac{f^2}{2}
 \int d^3 x_3 d^3 x_4 \langle  J_s  ( x_1 ; \epsilon_1) \mathcal{O} ( x_3) \mathcal{O} ( x_4) \rangle _0  \langle J_s ( x_2 ;  \epsilon_2)   \mathcal{O} (x_3) \mathcal{O}( x_4) \rangle_0 \, .
\end{align}
Here we have use the large $N$ factorization, where an $n$ point function is proportional to $N^{1-n/2}$ if it is normalized by two point functions.
The first and second types of contribution correspond to the left and right Witten diagrams in fig. \ref{critical0}, respectively.
There are also contributions at the higher order in $f$ and the corresponding Witten diagrams can be found in fig. \ref{pert}. The left diagram in fig. \ref{pert}  comes from the Witten diagram where the scalar propagates along the upper line of the loop in fig. \ref{2pt} and a higher spin field (or the scalar field) propagates along the lower line.  The right diagram in fig. \ref{pert} comes from the diagram with the scalar propagating along both lines.

\subsection{Anomalous dimensions at the IR fixed point}

\begin{figure}
  \centering
  \includegraphics[width=10cm]{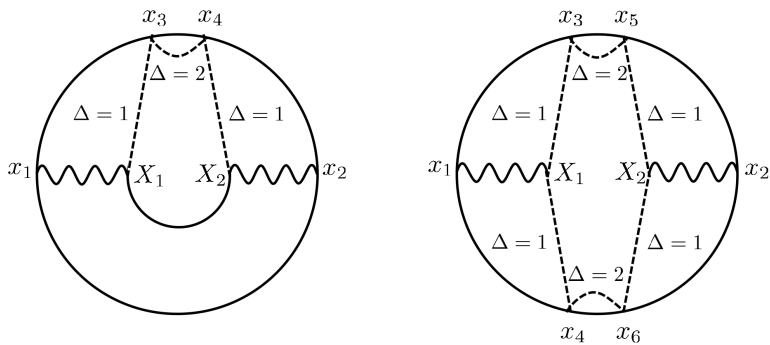}
  \caption{The effective Witten diagrams corresponding to the integrals we need to compute. The dual scaling dimensions of bulk scalar propagating along the lines are added.}
  \label{critical2}
\end{figure}

In order to compute the two point function at the IR limit, we first sum over the higher order contributions in  $f$ and then take the limit $f \to \infty$.
Let us first consider the integral $\tilde I_1$ in \eqref{type1}, which corresponds to the left diagram in fig. \ref{critical0}. The two point function $\langle \mathcal{O}(x_3) \mathcal{O}(x_4) \rangle_0$ in the integral corresponds to the dotted line between $x_3$ and $x_4$ in the left diagram of fig. \ref{critical0}.
At the higher order in $f$,
the scalar propagator receives corrections from boundary operator insertions as in the left diagram of fig. \ref{pert}.
After summing over the higher order corrections, 
the two point function is replaced by $\langle \mathcal{O}(x_3) \mathcal{O}(x_4) \rangle_f$
at the leading order in $1/N$.
Thus an integral we have to compute is 
\begin{align}
\label{ttype1}
 I_1 = \frac{f^2}{2}
\int d^3 x_3 d^3 x_4 \langle  J_s  ( x_1 ;  \epsilon_1)  J_s ( x_2 ;  \epsilon_2)   \mathcal{O} ( x_3) \mathcal{O} ( x_4)  \rangle _0 \langle  \mathcal{O}( x_3) \mathcal{O}( x_4)\rangle_f \, .
\end{align}
At $f \to \infty$, the two point function behaves as \eqref{sum1}.
The contribution from the contact term becomes the same integral as that at the first order in $f$,
and we have already seen that there is no contribution proportional to $\log |x_{12}|$.
Therefore, we can use 
\begin{align}
\langle  \mathcal{O}( x_3) \mathcal{O}( x_4)\rangle_f
 \sim \frac{1}{f^2} \frac{1}{4 \pi^4 N} \frac{1}{|x_{34}|^4} \, ,
\end{align}
and the $f^{-2}$ factor cancels the $f^2$ factor in \eqref{ttype1}.
The corresponding Witten diagram can be expressed as in the left diagram of fig. \ref{critical2}.

The sum over higher order corrections for the other integral $\tilde I_2$ in \eqref{type2} can be analyzed in a similar way.
Another integral we need to compute turns out to be
\begin{align}
\label{ttype2}
 I_2 = \frac{f^2}{2}
 \int d^3 x_3 d^3 x_4 d^3 x_5 d^3 x_6&  \langle  J_s  ( x_1 ; \epsilon_1) \mathcal{O} ( x_3) \mathcal{O} ( x_4) \rangle _0 G (x_{35})_f G(x_{46})_f \\ & \qquad  \times 
  \langle J_s ( x_2 ;  \epsilon_2)   \mathcal{O} (x_5) \mathcal{O}( x_6) \rangle_0 \, ,
  \nonumber
\end{align}
where $G (x)_f$ is defined in \eqref{sum2}. Notice that the delta function in $G (x)_f$ is necessary in order to incorporate the $f^2$ order contribution $\tilde I_2$ in \eqref{type2}.
Since $G(x)_f \sim f^{-1}$ for $f \to \infty$ as in  \eqref{sum2}, the factor cancels $f^2$ in front of the integral in \eqref{ttype2}.
The corresponding Witten diagram can be given as in the right diagram of fig. \ref{critical2}.

In this way, we have shown that the Witten diagram for the loop correction in fig. \ref{2pt} with the $\Delta = 2$ scalar boundary condition can be examined in terms of the products of tree level diagrams as in fig.\ref{critical2}. See appendix \ref{AppB} for more details.
This result is consistent with the previous one in \cite{Hartman:2006dy,Giombi:2011ya}, which was written with the momentum basis. 
In this sense we have elaborated their result by using the conformal perturbation theory such as to be suitable for our explicit computation.

In the rest of this paper, we compute the contributions proportional to $\log |x_{12}|$ in the two integrals $I_1$ \eqref{ttype1} and  $I_2$ \eqref{ttype2} at the limit of $f \to \infty$.
The results are summarized as 
\begin{align}
&I_1^{(1)} \sim - \frac{16}{ 3 \pi^2 } \frac{(2s)!}{(s!)^2}\frac{(x_{12}^-)^{2s}}{|x_{12}|^{4s +2}}
\log |x_{12}| \, , \label{I11} \\
&I_1^{(2)} \sim \frac{16}{\pi^2 } \frac{1}{ (2s-1)(2s + 1)}  \frac{(2s)!}{(s!)^2}  
 \frac{(x_{12}^-)^{2s}}{|x_{12}|^{4s + 2}} \log |x_{12}|  \label{I12}
\end{align}
with 
\begin{align}
I_1 = I_1^{(1)} + I_1^{(2)} 
\end{align}
and 
\begin{align}
 I_2 \sim \frac{32}{\pi^2 } \frac{s}{(2s - 1) (2s+1)} \frac{(2s)!}{(s!)^2}   \frac{(x_{12}^-)^{2s}}{|x_{12}|^{2 + 4s} } \log |x_{12}| \, .
 \label{I2}
\end{align}
Thus the sum over all contributions is
\begin{align}
 I_1^{(1)} + I_1^{(2)} +  I_2 
  \sim -  \frac{32 (s-2)}{3 \pi^2 (2s - 1) } \frac{(2s)!}{(s!)^2}   \frac{(x_{12}^-)^{2s}}{|x_{12}|^{2 + 4s} } \log |x_{12}| \, .
\end{align}
Comparing the expression in \eqref{anomalous}, we obtain
\begin{align}
\tau_s = \frac{16 (s-2)}{3 \pi^2 N (2s - 1)} \, , 
\end{align}
which reproduces \eqref{Ruhlr}.

As shown in \cite{Girardello:2002pp}, the Goldstone modes are the bound states of higher spin field and scalar field with $\Delta = 2$ as in \eqref{bound}, so it is expected that non-zero contributions arise only from the left diagram in fig. \ref{critical2}. However, this is not the case as seen in \eqref{I2}.
This can be explained as follows.
From the left diagram in fig. \ref{critical2}, there could be extra contributions, where a scalar propagates along the solid line between the bulk points $X_1$ and $X_2$.
Thus we  would have non-zero contributions from the right diagram, which cancel those from the left diagram. It would be nice to confirm this explicitly by computing the Witten diagrams in terms of bulk propagators.

\section{Details of computation}
\label{computation}

In this section, we derive the results in \eqref{I11}, \eqref{I12} and \eqref{I2}.
We start from the simplest case and then move to more involved ones.

\subsection{Integral $ I_1$}

Let us first consider the integral $ I_1$ in \eqref{ttype1}.
The integrand includes a four point function, which can be written
as a sum of two terms as
\begin{align}
\langle J_s ( x_1 ;  \epsilon_1 ) J_s ( x_2 ;  \epsilon_2) \mathcal{O} ( x_3) \mathcal{O} ( x_4)  \rangle _0 = K_1 ( x_i ;  \epsilon_i) + K_2 ( x_i ;  \epsilon_i) \, ,
\end{align}
where
\begin{align}
\label{tL1}
K_1 ( x_i ; \epsilon_i) &=
16 N \frac{1}{|x_{34}|} \frac{1}{|x_{41}|} 
e^{ - \epsilon_1 \cdot \overleftarrow{\partial}_1 } 
 \cos \left( 2 \sqrt{ \epsilon_1 \cdot  \overleftarrow{\partial}_1 \epsilon_1 \cdot  \overrightarrow{\partial}_1}\right) 
e^{  \epsilon_1 \cdot \overrightarrow{\partial}_1 } 
 \frac{1}{|x_{12}|}  \\ &\times
 e^{ - \epsilon_2 \cdot \overleftarrow{\partial}_2 } 
 \left.  \cos \left( 2 \sqrt{ \epsilon_2 \cdot  \overleftarrow{\partial}_2 \epsilon_2 \cdot  \overrightarrow{\partial}_2}\right) 
 e^{  \epsilon_2 \cdot \overrightarrow{\partial}_2 } 
 \frac{1}{|x_{23}|} \right|_{\epsilon_1^s \epsilon_2^s}  + (3 \leftrightarrow 4) \nonumber
\end{align}
and
\begin{align}
\label{tL2}
K_2 ( x_i ; \epsilon_i) & =
16 N  \frac{1}{|x_{41}|} 
e^{ - \epsilon_1 \cdot \overleftarrow{\partial}_1 } 
 \cos \left( 2 \sqrt{ \epsilon_1 \cdot  \overleftarrow{\partial}_1 \epsilon_1 \cdot  \overrightarrow{\partial}_1}\right) 
e^{  \epsilon_1 \cdot \overrightarrow{\partial}_1 } 
 \frac{1}{|x_{13}|}  \\ & \times  \frac{1}{|x_{32}|}
 e^{ - \epsilon_2 \cdot \overleftarrow{\partial}_2 } 
 \left.  \cos \left( 2 \sqrt{ \epsilon_2 \cdot  \overleftarrow{\partial}_2 \epsilon_2 \cdot  \overrightarrow{\partial}_2}\right) 
 e^{  \epsilon_2 \cdot \overrightarrow{\partial}_2 } 
 \frac{1}{|x_{24}|} \right|_{\epsilon_1^s \epsilon_2^s}  \nonumber \, .
\end{align}
Here we have used the $n$-point correlator of the generating function in \eqref{npt}.

It is convenient to separate the integral $I_1$ following the expression of four point function as 
\begin{align}
 I_1 = I_1^{(1)} + I_1^{(2)} \, , \quad
 I_1^{(a)} = \int d^3 x_3 d^3 x_4   L_a (x_i ; \epsilon_i) \, ,
\end{align}
where
\begin{align}
\label{L1}
 L_1 ( x_i ; \epsilon_i) &=
8 N f^2 \frac{1}{|x_{34}|} \frac{1}{|x_{41}|} 
e^{ - \epsilon_1 \cdot \overleftarrow{\partial}_1 } 
 \cos \left( 2 \sqrt{ \epsilon_1 \cdot  \overleftarrow{\partial}_1 \epsilon_1 \cdot  \overrightarrow{\partial}_1}\right) 
e^{  \epsilon_1 \cdot \overrightarrow{\partial}_1 } 
 \frac{1}{|x_{12}|}  \\ &\times
 e^{ - \epsilon_2 \cdot \overleftarrow{\partial}_2 } 
 \left.  \cos \left( 2 \sqrt{ \epsilon_2 \cdot  \overleftarrow{\partial}_2 \epsilon_2 \cdot  \overrightarrow{\partial}_2}\right) 
 e^{  \epsilon_2 \cdot \overrightarrow{\partial}_2 } 
 \frac{1}{|x_{23}|} \right|_{\epsilon_1^s \epsilon_2^s} \frac{\tilde C}{|x_{43}|^4} 
  + (3 \leftrightarrow 4)\nonumber 
\end{align}
and
\begin{align}
\label{L2}
 L_2 ( x_i ;  \epsilon_i) & =
8 N f^2 \frac{1}{|x_{41}|} 
e^{ - \epsilon_1 \cdot \overleftarrow{\partial}_1 } 
 \cos \left( 2 \sqrt{ \epsilon_1 \cdot  \overleftarrow{\partial}_1 \epsilon_1 \cdot  \overrightarrow{\partial}_1}\right) 
e^{  \epsilon_1 \cdot \overrightarrow{\partial}_1 } 
 \frac{1}{|x_{13}|}  \\ & \times  \frac{1}{|x_{32}|}
 e^{ - \epsilon_2 \cdot \overleftarrow{\partial}_2 } 
 \left.  \cos \left( 2 \sqrt{ \epsilon_2 \cdot  \overleftarrow{\partial}_2 \epsilon_2 \cdot  \overrightarrow{\partial}_2}\right) 
 e^{  \epsilon_2 \cdot \overrightarrow{\partial}_2 } 
 \frac{1}{|x_{24}|} \right|_{\epsilon_1^s \epsilon_2^s} \frac{\tilde C}{|x_{43}|^4} 
 \nonumber 
\end{align}
for $f \to \infty$.
The coefficient is $\tilde C = 1/(4 \pi^4 f^2 N)$ as in \eqref{sum1}. 
In the following we examine the integrals $I_1^{(1)}$ and $I_1^{(2)}$ separately.

\subsubsection{Integral $I_1^{(1)}$}

In order to compute the integral $I_1^{(1)}$, we need to pick up the term proportional to 
$\epsilon_1^s \epsilon_2^s$ in \eqref{L1}.
This can be done as in (4.108) of \cite{Giombi:2009wh}
\begin{align}
&  L_1 ( x_i ; \epsilon_i) = 8 N f^2 \tilde C \left( \frac{(2s)!}{s!}\right)^2
\sum_{n,m=0}^s \frac{(-1)^{n+m}}{(2n)!(2m)!(2s-2n)!(2s-2m)!} \\
&\times \left[ (\epsilon_1 \cdot \partial_1)^{s - n} \frac{1}{|x_{41}|}\right]
 \left[ (\epsilon_2 \cdot \partial_2)^{s - m } \frac{1}{|x_{23}|}\right]
 \left[(\epsilon_1 \cdot \partial_1)^{n} (\epsilon_2 \cdot \partial_2)^{ m } \frac{1}{|x_{12}|}\right] \frac{1}{|x_{34}|^5} + (3 \leftrightarrow 4) \, ,
 \nonumber
\end{align}
where \eqref{formula1} has been applied.
Therefore, we can obtain the value of $ I_1^{(1)}$, once we can evaluate the integral
\begin{align}
 P_1 = \int d ^3x_3 d ^3x_4 \frac{1}{|x_{14} | |x_{43} |^5 |x_{32}| } \, .
 \label{P1_0}
\end{align}
This is because $ I_1^{(1)}$ could be given in terms of derivatives 
with respect to $x_1 , x_2$. 
However, the integral diverges if we naively apply the formula \eqref{vertex}.
In the following we shall develop a way to regularize the divergence by applying the dimensional regularization.
Using the regularization, we will find out the contribution proportional to $\log |x_{12}|$.

We would like to compute the integral $P_1$ with the momentum basis. However, it is not possible to do so for $|x_{34}|^{-5}$ since the coefficient in \eqref{formula} diverges. 
To avoid this problem,
we rewrite $|x_{34}|^{-5} = |x_{34}|^{-2 t} |x_{34}|^{-5 + 2 t}$ with $t$ not half-integer.
The result should not depend on the choice of $t$, but we keep $t$ generic to make the independence manifest.
This way of expression leads to
\begin{align}
 P_1 &= \int d ^3x_3 d ^3x_4 \frac{1}{|x_{14} | |x_{34}|^{2t} |x_{34}|^{5-2t}  |x_{32}| } \\
 & = 2^{-7} \pi^{-6} (a(\tfrac12))^2 a(t) a(\tfrac52 - t)\int  d ^3x_3 d ^3x_4  \prod_{i=1}^4 d ^3 k_i \frac{e^{i k_1 \cdot x_{14} + i (k_2 + k_3) \cdot x_{34} + i k_4 \cdot x_{32} }  }{|k_1|^2 |k_2|^{3 - 2 t} |k_3|^{2 t- 2} |k_4|^{2}} \nonumber \, .
\end{align}
The integration over $x_3,x_4$ yields the product of delta function as $(2 \pi)^6 \delta^{(3)}(k_1 + k_2 + k_3) \delta^{(3)} (k_2 + k_3 + k_4)$. 
Thus we obtain
\begin{align}
\label{P1}
 &P_1  = \frac{1}{2 \pi} \int  d ^3 k_1 \frac{e^{i k_1 \cdot x_{12}  } }{|k_1|^4} F_1 (k_1) \, , \\
  &F_1 (k_1) =a(t) a(\tfrac52 - t)
 \int  d ^3 k_2   \frac{1 }{ |k_2|^{3 - 2 t} |k_1 + k_2|^{2t - 2} }  
 \label{F1}
\end{align}
after the integration over $k_3,k_4$.

The integral over $k_2$ in $F_1$ diverges, so we would like to apply the dimensional regularization here.
Introducing the Feynman parameter as
\begin{align}
 \frac{1}{A^{m_1}_1 A^{m_2}_2 \cdots A^{m_n}_n}
  = \int_0^1 d y_1 \cdots d y_n \delta (\sum y_i - 1)  \frac{\prod y_i^{m_i - 1}}{(\sum y_i A_i)^{\sum m_i}} \frac{\Gamma (m_1 + \cdots + m_n)}{\Gamma (m_1 ) \cdots \Gamma (m_n)} \, ,
  \label{Feynman}
\end{align}
we can rewrite $F_1$ as
\begin{align}
F_1 (k_1) &= \frac{\Gamma(\frac12)}{\Gamma ( t) \Gamma (\frac52 - t)}                                                                            
 \int  d ^3 k_2   \int_0^1 d y \frac{ (1  - y)^{1/2 - t } y^{t - 2} }
 { (( 1 - y) |k_2|^2 + y|k_1 + k_2|^2)^{1/2}}  \\
 &= \frac{\Gamma(\frac12)}{\Gamma ( t) \Gamma (\frac52 - t)}                                                                            
  \int_0^1 d y  \int  d ^3 k_2  \frac{ (1  - y)^{1/2 - t } y^{t - 2} }
  { ( |k_2 + y k_1 |^2 + y(1 - y) |k_1 |^2)^{1/2}}  \, . \nonumber 
\end{align}
Using 
\begin{align}
\int \frac{d^d \ell}{(2 \pi)^d} \frac{1}{(\ell^2 + \Lambda)^n}
 = \frac{1}{(4 \pi)^{d/2}} \frac{\Gamma (n - \frac{d}{2})}{\Gamma (n)} \left( \frac{1}{\Lambda}\right)^{n - \frac{d}{2}} \, ,
 \label{dr}
\end{align}
we have
\begin{align}
\int \frac{d^d \ell}{(2 \pi)^d} \frac{1}{(\ell^2 + \Lambda)^{1/2}}
 \to  \frac{1}{ 2 (2 \pi) ^2 } \Lambda \left( - \frac{2}{\epsilon} - 1 + \gamma + \log \Lambda - \log 4 \pi + \mathcal{O} (\epsilon) \right) 
\end{align}
for $d = 3 - \epsilon$.
Since we are interested in the contribution proportional to $\log |x_{12}|$, we keep the part which produces such terms. Thus we keep
\begin{align}
F_1 (k_1) \sim \frac{\Gamma(\frac12) (2 \pi)^3}{\Gamma ( t) \Gamma (\frac52 - t)}                                                                            
 \frac{|k_1|^2}{2 (2 \pi)^2} \log |k_1|^2  \int_0^1 dy \, y^{t-1} (1 - y)^{3/2 -t} 
 = \frac{4}{3} \pi |k_1|^2 \log |k_1|^2 \, ,
\end{align}
which leads to
\begin{align}
P_1 \sim \frac{4}{ 3} \int d^3 k_1 \frac{e^{i k_1 \cdot x_{12}}}{|k_1|^2} \log |k_1| 
      \sim - \frac{ 8 \pi^2}{3}  |x_{12}|^{-1}  \log |x_{12}| \, .
\end{align}
Here we have used \eqref{P1} and the formula \eqref{Flog}.

Now we can obtain the expression of $I_1^{(1)}$ using the above result.
For $\epsilon_1 = \epsilon_2$, we find
\begin{align}
I_1^{(1)} & \sim - 2 \cdot \frac{8 \pi^2}{3} \cdot 8 N f^2 \tilde C \left( \frac{(2s)!}{s!}\right)^2
\sum_{n,m=0}^s \frac{(-1)^{n+m}}{(2n)!(2m)!(2s-2n)!(2s-2m)!} \\
&\times 
 \left[ (\epsilon_1 \cdot \partial_2)^{2 s - n - m } \frac{1}{|x_{12}|}\right]
 \left[ (\epsilon_1 \cdot \partial_2)^{ n + m } \frac{1}{|x_{12}|}\right] \log |x_{12}|  \nonumber\\
 &= - \frac{128 \pi^2 N f^2 \tilde C }{3}  \left( \frac{(2s)!}{s!}\right)^2
 \sum_{n,m=0}^s \frac{(-1)^{n+m}}{(2n)!(2m)!(2s-2n)!(2s-2m)!} \nonumber \\
 &\times \pi^{-1}
 \Gamma(2s - n - m + \tfrac12 ) \Gamma (n+m+\tfrac12)\frac{(x_{12}^-)^{2s}}{|x_{12}|^{4s +2}}
\log |x_{12}|  \, .
 \nonumber
\end{align}
Here we have used a convenient formula
  \begin{align}
   (\epsilon_1 \cdot \partial_2)^a \frac{1}{|x_{12}|^b} = \frac{\Gamma(a + \frac{b}{2})}{\Gamma (\frac{b}{2})} \frac{(x_{12}^-)^{a}}{|x_{12}|^{2a+b} } \, .
   \label{convenient}
  \end{align}
Applying the formula \eqref{formula15} to the sum over $n,m$, we arrive at 
\begin{align}
I_1^{(1)} & \sim - \frac{128 \pi^2 N f^2 \tilde C }{3} \left( \frac{(2s)!}{s!}\right)^2 \frac{1}{2 (2s)!}\frac{(x_{12}^-)^{2s}}{|x_{12}|^{4s +2}}
\log |x_{12}|  \\
&= - \frac{16}{ 3 \pi^2 } \frac{(2s)!}{(s!)^2}\frac{(x_{12}^-)^{2s}}{|x_{12}|^{4s +2}}
\log |x_{12}| \nonumber \, ,
\end{align}
where we have used $\tilde C = 1/(4 \pi^4 f^2 N)$.
In this way we have obtained the result in \eqref{I11}.

\subsubsection{Integral $I_1^{(2)}$}

Let us move to the integral $I_1^{(2)}$.
As for $I_1^{(1)}$, we would like to pick up the term proportional to $\epsilon_1^s \epsilon_2^s$ in $L_2$ given in  \eqref{L2}.
For the purpose it is useful to utilize  the following three point function, which was examined in (4.103) of \cite{Giombi:2009wh} as
\begin{align}
\nonumber
  &\langle  J_s (x_1 ; \epsilon_1) \mathcal{O} ( x_3) \mathcal{O} (x_4) \rangle _0 
= \left. 8 N \frac{1}{|x_{34}|} \frac{1}{|x_{41}|} e^{\overleftarrow{\partial}_{+,1}} 
  \cos \left[ 2 \sqrt{ \overleftarrow{\partial}_{+,1} \overrightarrow{\partial}_{+,1} }\right] e^{ - \overrightarrow{\partial}_{+,1}} \frac{1}{|x_{13}|} \right|_{\epsilon_1^s} \\ 
  & \quad = 8N \left( \frac{(2s)!}{s!}\right)  \frac{1}{|x_{34}|}
  \sum_{n=0}^s \frac{(-1)^n}{(2n)!(2s-2n)!} \left[ \partial_{+,1}^{s-n} \frac{1}{|x_{13}|}  \right] \left[ \partial_{+,1}^{n} \frac{1}{|x_{41}|}  \right] 
  \label{GY3pt}
 \end{align}
 with $\partial_{+,i} = \epsilon_1 \cdot \partial_{x_i}$.
%  Here the formula \eqref{formula1} is again useful.
With the help of this expression,
we can rewrite the integral $I_1^{(2)}$ as
\begin{align}
 I_1^{(2)}  =  8 N f^2 \tilde C \left( \frac{(2s)!}{s!}\right)^2  
 \sum_{n,m=0}^s
 \frac{(-1)^{n+m} }{(2n)! (2s-2n)! (2m)! (2s-2m)!} B_{m,n} \, ,
\end{align}
where
\begin{align}
\label{Bmn}
B_{m,n} =   \int d^3 x_3 d^3 x_4  \left[\partial_{+,1}^{n} \frac{1}{ |x_{31} | } \right]\left[ \partial_{+,2}^{m} \frac{1}{|x_{24}|}\right]   \left[ \partial_{+,1}^{s-n} \frac{1}{|x_{14}|}  \right]\left[ \partial_{+,2}^{s-m} \frac{1}{|x_{32}|} \right] \frac{1}{|x_{34}|^4} 
\nonumber
\end{align}
for $\epsilon_1 = \epsilon_2$.

We need to perform the integration over $x_3,x_4$.
The integration over $x_4$ can be done as
\begin{align}
&\partial_{+,2}^m \partial_{+,1}^{s-n} 
\int d ^3 x_4 \frac{1}{ |x_{14}| |x_{24}| |x_{34}|^4} 
 = v(\tfrac12,\tfrac12,2) \partial_{+,2}^m \partial_{+,1}^{s-n} 
 \frac{1}{|x_{23}|^2 |x_{31}|^2 |x_{12}|^{-1}} \\
 & \quad
 = - 2 \pi \sum_{k = 0}^{m} \sum_{l = 0}^{s-n} 
 \binom{m}{k} \binom{s-n}{l}
\left[ \partial_{+,2}^k  \frac{1}{|x_{23}|^2 } \right]
\left[ \partial_{+,1}^l  \frac{1}{|x_{31}|^2 } \right]
\left[ \partial_{+,2}^{m-k}   \partial_{+,1}^{s-n-l} |x_{12}|\right]
\nonumber 
\end{align}
by applying the formula \eqref{vertex}.
Using  \eqref{convenient},
we can rewrite
\begin{align}
\label{del1}
& \left[\partial^{s-m}_{+,2}\frac{1}{|x_{32}|} \right] \left[ \partial^k_{+,2} \frac{1}{|x_{32}|^2} \right]
 = \frac12 \frac{\Gamma (s - m + \frac12) \Gamma (k+1)}{\Gamma (s - m + k + \frac32)}
 \partial^{s-m + k}_{+,2} \frac{1}{|x_{32}|^3}  \, , \\
& \left[\partial^{n}_{+,1} \frac{1}{|x_{31}|} \right] \left[ \partial^l_{+,1} \frac{1}{|x_{31}|^2}
\right]  = \frac12 \frac{\Gamma (n + \frac12) \Gamma (l+1)}{\Gamma (n+l + \frac32)}
 \partial^{n+l}_{+,1} \frac{1}{|x_{31}|^3} \,  .
 \label{del2}
\end{align}
These expressions imply that the integral $I_1^{(2)}$ reduces to the derivatives of 
\begin{align}
P_2 = \int d^3 x_3 \frac{1}{|x_{32}|^3 |x_{31}|^3} 
\label{P2_0}
\end{align}
with respect to $x_1,x_2$. However, the integration over $x_3$ diverges and a regularization is needed as for $I_1^{(1)}$.

We would like to compute the integral $P_2$  \eqref{P2_0} with the momentum basis.
Since we cannot apply the formula \eqref{formula} due to $\Gamma (0)$
in the coefficient, we again set as $|x|^{3} = |x|^{2t} |x|^{3-2t}$ ($t$ is not half-integer) and perform the Fourier transforms to them separately. 
Thus we rewrite the integral as
\begin{align}
P_2 &= \int d^3 x_3 \frac{1}{|x_{32}|^{2t}  |x_{32}|^{3-2t}  |x_{31}|^{2u} |x_{31}|^{3-2u}} \\
&= \frac{1}{(2 \pi)^6} \int d^3 x_3 \int \prod_{i=1}^4 d ^3 k_i \frac{e^{i (k_1 + k_2) \cdot  x_{32} + i (k_3 + k_4) \cdot  x_{31}}}{|k_1|^{3-2t} |k_2|^{2t}  |k_3|^{3-2u}  |k_4|^{2u} } \, .
\nonumber  
\end{align}
The integration over $x_3$ yields $(2 \pi)^3 \delta^{(3)} (\sum k_i)$, and after shifting $k_1 \to k_1 - k_2$ we have
\begin{align}
\label{P2}
P_2 
&= \frac{1}{(2 \pi)^3} \int \prod_{i=1}^3 d ^3 k_i \frac{e^{i k_1 \cdot  x_{21}}}{|k_1 - k_2 |^{3-2t} |k_2|^{2t}  |k_1 + k_4|^{3-2u}  |k_4|^{2u} } \\
&= \frac{1}{(2 \pi)^3} \int d ^3 k_1 e^{ - i k_1 \cdot  x_{12}} F_2 (t , k_1) F_2 (u , k_1) \, ,
\nonumber  
\end{align}
where
\begin{align}
F_2 (t,k_1) = \int d ^3 k_2 \frac{1}{|k_1 - k_2 |^{3-2t} |k_2|^{2t} } \, .
\end{align}
The task now is to pick up the part producing contributions proportional to $\log |x_{12}|$.

Introducing the Feynman parameter \eqref{Feynman}, the integral becomes
\begin{align}
F_2 (t , k_1) = \frac{\Gamma (\frac32)}{ \Gamma (\frac32 - t) \Gamma (t)}\int d ^3 k_2 \int_0^1 dy \frac{y^{1/2 - t} (1-y)^{ t- 1}}
{ ( y|k_1 - k_2 |^2 + (1-y) |k_2|^2 )^{3/2}} \\
 = \frac{\Gamma (\frac32)}{ \Gamma (\frac32 - t) \Gamma (t)} \int_0^1 d y \int d ^3 k_2
 \frac{y^{1/2 - t } (1-y)^{t- 1}}{ ( |k_2 - y k_1|^2  + y  (1-y) |k_1|^2 )^{3/2}}\, .
 \nonumber
\end{align}
From \eqref{dr}, the dimensional regularization gives
\begin{align}
\int \frac{d ^d \ell}{(2 \pi)^d} \frac{1}{( \ell^2 + \Lambda ) ^{3/2}}
 \to \frac{1}{(2 \pi)^2} \left( \frac{2}{\epsilon} - \log \Lambda - \gamma + \log 4 \pi + \mathcal{O}(\epsilon) \right)
\end{align}
for $d = 3 - \epsilon$. The term proportional to $\log |k_1|$ becomes
\begin{align}
F_2 (t , k_1) \sim - 4 \pi  \log |k_1| \, .
\end{align}
Thus the term proportional to $\log |x_{12}|$ in $P_2$ \eqref{P2} becomes%
\footnote{The integral $P_2$ includes also the term proportional to 
$ \int d^3 k_1 e^{- i k_1 \cdot x_{12}} \log |k_1| $, but we can see from \eqref{Flog0} that it gives no contribution proportional to $\log |x_{12}|$.}
\begin{align}
P_2 \sim \frac{2}{\pi} \int d^3 k_1 e^{- i k_1 \cdot x_{12}} ( \log |k_1| )^2
 \sim \frac{8 \pi}{|x_{12}|^3} \log |x_{12}| \, ,
 \label{P2log}
\end{align}
where we have used the formula \eqref{Flog2}.

Using the expression of $B_{m,n}$ in \eqref{Bmn} with \eqref{del1} and \eqref{del2}, we find 
\begin{align}
\nonumber
 B_{m,n}  \sim &- 16 \pi^{2} \sum_{k=0}^m
 \sum_{l = 0}^{s-n} 
 \binom{m}{k} \binom{s-n}{l} 
 \frac14 \frac{\Gamma (s- m + \frac12) \Gamma (k+1) \Gamma (n + \frac12) \Gamma (l+1)}{\Gamma (s - m + k + \frac32) \Gamma (n + l + \frac32)} \\ &\times
 \left[  \partial_{+,2}^{s+m-n -k -l} |x_{12}|\right]
 \left[ \partial_{+,2}^{s-m + n +k +l} \frac{1}{|x_{12}|^3} \right]  \log |x_{12}|\\
  = & 4 \pi  \sum_{k=0}^m
  \sum_{l = 0}^{s-n} 
  \binom{m}{k} \binom{s-n}{l} 
  \frac{\Gamma (s- m + \frac12) \Gamma (k+1) \Gamma ( n + \frac12) \Gamma (l+1)}{\Gamma (s - m + k + \frac32) \Gamma ( n + l + \frac32)} \nonumber \\ &  \times
   \Gamma (s + m - n - k - l - \tfrac12) \Gamma (s - m + n + k + l  + \tfrac32) \frac{(x_{12}^-)^{2s}}{|x_{12}|^{4s + 2}} \log |x_{12}|  \, .\nonumber 
 \end{align}
 We have used \eqref{convenient} as well.
 Then the formula \eqref{formula7} leads to
\begin{align}
I_1^{(2)} & \sim 8 N f^2 \tilde C \left( \frac{(2s)!}{s!}\right)^2  
 \frac{8 \pi^2}{(2s-1)(2s + 1)!} \frac{(x_{12}^-)^{2s}}{|x_{12}|^{4s + 2}} \log |x_{12}| \\
 & = \frac{16}{\pi^2  (2s-1)(2s + 1)}  \frac{(2s)!}{(s!)^2}  
 \frac{(x_{12}^-)^{2s}}{|x_{12}|^{4s + 2}} \log |x_{12}| 
   \nonumber 
\end{align}
as in \eqref{I12}.

\subsection{Integral $I_2$}

Finally we examine the integral $I_2$ in \eqref{ttype2}. 
Using the expression of three point function in \eqref{GY3pt}, the integral can be written as
 \begin{align}
 \nonumber
 & I_2 = \frac{f^2}{2} (8N)^2  \left( \frac{(2s)!}{s!}\right)^2  
  \sum_{n,m=0}^s
  \frac{(-1)^{n+m}}{(2n)! (2s-2n)! (2m)! (2s-2m)!} 
  \int d^3 x_3 d^3 x_4  d^3 x_5 d^3 x_6 \\
   & \qquad   \left[ \partial_{+,1}^{n} \frac{1}{|x_{31}|} \right] \left[ \partial_{+,1}^{s-n} \frac{1}{|x_{14}|}  \right] \left[ \partial_{+,2}^{m} \frac{1}{|x_{25}|}\right] \left[ \partial_{+,2}^{s-m} \frac{1}{|x_{62}|} \right] \frac{1}{|x_{34}|}  \frac{1}{|x_{56}|} \frac{\tilde C '}{|x_{45}|^4  }   \frac{\tilde C '}{|x_{36}|^4  } \, . \nonumber
 \end{align}
Here  $\tilde C ' = - 1/(4 \pi^4 f N)$, which comes from \eqref{sum2}.

In this case we have four integral variables $x_3$, $x_4$, $x_5$ and $x_6$.
We can integrate over  $x_4$ and $x_6$ by applying the formula \eqref{vertex} as 
\begin{align}
 \partial_{+,1}^{s-n} \int d ^3 x_4 \frac{1}{|x_{14}| |x_{34}| |x_{45}|^4} 
  &= - 2 \pi \partial_{+,1}^{s-n} \frac{1}{| x_{35}|^2 |x_{15}|^2 |x_{13}|^{-1} } \\
  & = - 2 \pi \sum_{l=0}^{s-n} \binom{s-n}{l} \frac{1}{|x_{35}|^2}
       \left[ \partial_{+,1}^{s - n - l} \frac{1}{|x_{15}|^2} \right]    \left[ \partial_{+,1}^{l} 
           |x_{13}|\right]   \, , \nonumber \\
\partial_{+,2}^{s-m} \int d ^3 x_6 \frac{1}{|x_{62}| |x_{56}| |x_{36}|^4} 
 &= - 2 \pi \partial_{+,2}^{s-m} \frac{1}{| x_{35}|^2 |x_{23}|^2 |x_{25}|^{-1}}  \\
  &= - 2 \pi \sum_{k=0}^{s-m} \binom{s-m}{k} \frac{1}{|x_{35}|^2}
     \left[ \partial_{+,2}^{s - m - k} \frac{1}{|x_{23}|^2} \right]    \left[ \partial_{+,2}^{k} 
         |x_{25}|\right]   \, . \nonumber 
\end{align}

In order to integrate over $x_3$, we need to collect the terms involving $|x_{31}|$.
For $n +  l \neq 0$, we can rewrite them as 
\begin{align}
\partial_{+,1}^n \frac{1}{|x_{31}|} \partial_{+,1}^l |x_{31}|
 = \frac{1}{\pi} \frac{ \Gamma (n + \tfrac12) \Gamma (l - \tfrac12) }{\Gamma (n + l)}
 \partial^{n+l}_{+,1} \log |x_{31}|
 \label{nlnzero}
\end{align}
by applying \eqref{convenient}.
For $n = l = 0$ we can easily see that this does not hold.
Using \eqref{nlnzero} the integral over $x_3$ becomes
\begin{align}
& \frac{1}{\pi} \frac{ \Gamma (n + \tfrac12) \Gamma (l - \tfrac12) }{\Gamma (n + l)}
 \partial^{n+l}_{+,1} \partial_{+,2}^{s - m -k} \int d^3 x_3 \frac{\log|x_{31}|}{|x_{35}|^4 |x_{23}|^2 } \nonumber \\
& \quad  =  \frac{\pi^2}{2} \frac{ \Gamma (n + \tfrac12) \Gamma (l - \tfrac12) }{\Gamma (n + l)}
   \partial^{n+l}_{+,1} \partial_{+,2}^{s - m -k} \left( \frac{|x_{12}|}{|x_{52}|^3 |x_{15}| } \right)\nonumber \\
   & \quad  = 
  \frac{\pi^2}{2} \frac{ \Gamma (n + \tfrac12) \Gamma (l - \tfrac12) }{\Gamma (n + l)}
    \sum_{p=0}^{n+l} \sum_{q=0}^{s - m - k} \binom{n+l}{p} \binom{s-m-k}{q}
     \nonumber \\ & \quad  \times 
      \left [\partial^{n+l - p}_{+,1} \partial_{+,2}^{s - m -k - q}  |x_{12}| \right]
      \left[ \partial^{ p}_{+,1} \frac{1}{|x_{15}| } \right]
      \left[ \partial^q_{+ , 2} \frac{1}{|x_{52}|^3} \right]  \, .
\end{align}
Here the formula \eqref{vertexlog} has been utilized.
Then the integration over $x_5$ can be performed by rewriting as
\begin{align}
&\left[ \partial^p_{+,1} \frac{1}{|x_{51}|} \right]
\left[ \partial^{s - n - l}_{+,1} \frac{1}{|x_{51}|^2} \right]
 = \frac12 \frac{\Gamma (p + \frac12) \Gamma (s - n - l + 1)}{\Gamma (s - n - l + p + \frac32)}
 \partial_{+,1}^{s - n - l + p} \frac{1}{|x_{51}|^3} \, , \\
& \left[ \partial^m_{+,2} \frac{1}{|x_{52}|} \right]
  \left[ \partial^k_{+,2}  |x_{52}| \right]
   \left[ \partial^q_{+,2} \frac{1}{|x_{52}|^3} \right]
    = - \frac{1}{2 \pi} 
    \frac{\Gamma (m + \frac12) \Gamma (k - \frac12) \Gamma (q + \frac32)}{\Gamma (m + k +q + \frac32)} \partial_{+,2}^{m+k+q} \frac{1}{|x_{52}|^3} 
    \nonumber
\end{align}
and using \eqref{P2log}.

Let us suppose that \eqref{nlnzero} holds even for $n = l = 0$.
Then,  after the integration over $x_3$, $x_4$, $x_5$ and $x_6$,  the integral $I_2$ becomes
 \begin{align}
 \nonumber 
 & I_2  \sim  \frac{f^2}{2} (8N)^2  (\tilde C ' ) ^2 \left( \frac{(2s)!}{s!}\right)^2 (- 2 \pi)^2
 \frac{\pi^2}{2}  \left( -  \frac{1}{4 \pi} \right)  8 \pi 
 \frac{1}{\Gamma (- \frac12) \Gamma (\frac32)}
 \frac{(x_{12}^-)^{2s}}{|x_{12}|^{2 + 4s} } \log |x_{12}| \\
 & \qquad \qquad \qquad \times \sum _{m,n=0}^s \sum _{k=0}^{s-m} \sum _{l=0}^{s-n} \sum _{p=0}^{l+n} \sum _{q=0}^{s -k-m} H (s;n,m;k,l;p,q) \, ,
 \label{sumover}
 \end{align}
where $H (s;n,m;k,l;p,q)$ is defined in \eqref{Hf}.
Using the formula \eqref{formula8}, we find 
 \begin{align}
 \nonumber
  I_2 \sim \frac{32}{\pi^2 } \frac{s}{(2s - 1) (2s+1)} \frac{(2s)!}{(s!)^2}   \frac{(x_{12}^-)^{2s}}{|x_{12}|^{2 + 4s} } \log |x_{12}| \, ,
 \end{align}
which coincides with \eqref{I2}.
Here we have used $\tilde C ' = - 1/(4 \pi^4 f N)$.

The above result is the correct one since we can show that there is no contribution from $n = l=0$.
For $m + k \neq 0$, we can repeat the above computation by replacing
$(n,l , x_1 , x_3)$ with $(m,k,x_2 , x_5)$. As we see from \eqref{formula9},
the summation vanishes in this case.
For $m = k=0$, the integral becomes the derivative of
\begin{align}
\int d^3 x_3 d^3 x_5
\frac{1}{|x_{15}|^2 |x_{53}|^4 |x_{32}|^2} 
\end{align}
with respect to $x_1 , x_2$. Performing Fourier transforms, we find
\begin{align}
\int d^3 x_3 d^3 x_5 \prod_{i=1}^3 d^3 k_i
 \frac{e^{i  k_1 \cdot x_{15} + i k_2 \cdot x_{53} + i k_3 \cdot x_{32}}}{|k_1| |k_2|^{-1} |k_3|} \propto \int d^3 k_1  \frac{e^{ik_1 \cdot x_{12}}}{|k_1|} 
 \propto \frac{1}{|x_{12}|^2}\, ,
\end{align}
which means that there is no contribution proportional to $\log |x_{12}|$.
Therefore, we can neglect the case with $n = l=0$.
Owing to \eqref{formula9} and \eqref{formula10},
we can safely sum over all ranges of parameters in \eqref{sumover}
and apply the formula \eqref{formula8}.
In this way, we derive the result in \eqref{I2}.

\section{Conclusion}
\label{conclusion}

In this paper, we have examined the breaking of higher spin gauge symmetry in the 4d minimal bosonic Vasiliev theory \cite{Vasiliev:1995dn,Vasiliev:1999ba}, which is dual to the critical 3d O$(N)$ vector model \cite{Klebanov:2002ja}.
The dual CFT suggests that the symmetry breaking is due to the change of boundary condition for scalar field and the masses of higher spin fields come from  a loop effect.
The masses can be read off from the anomalous dimensions of dual currents, and they were obtained in  \cite{Ruhl:2004cf} as \eqref{Ruhlr} from the 3d critical model (see also \cite{Skvortsov:2015pea,Giombi:2016hkj}) at the leading order in $1/N$.
The anomalous dimensions can be calculated from the bulk theory using Witten diagrams as in fig.~\ref{2pt}.
We establish the relation between bulk Witten diagrams and boundary conformal perturbation theory using the fact that the shift of bulk scalar propagator with Dirichlet boundary condition can be represented by the insertions of boundary deformation operators \eqref{def}.
Reproducing the anomalous dimensions in the conformal perturbation theory, we provided an additional support for the bulk picture suggested by the dual CFT.

There are the following future problems;
It is desired to compute the Higgs masses from the one-loop corrections to the bulk higher spin propagator as was done in the spin 2 example \cite{Porrati:2001db,Duff:2004wh}.
In particular,  it would be nice if we can check that there is no contribution to the mass from the diagram, where the scalar field propagates along both sides of the loop in fig. \ref{2pt} as mentioned at the end of section \ref{methods}.
 Furthermore,
it would be useful to compare other methods to obtain  the anomalous dimensions from the boundary critical model as in \cite{Ruhl:2004cf,Maldacena:2012sf,Skvortsov:2015pea,Giombi:2016hkj}.%
\footnote{It would be also interesting to see the relation to the higher spin symmetry breaking examined in \cite{Leigh:2012mz}.}
We have used the conformal perturbation theory since it is directly connected to the computation with bulk Witten diagrams.
However, the method itself should be useful to compute the anomalous dimensions as well particularly for marginal deformations as in \cite{Creutzig:2015hta}.
It should be possible to work in generic dimensions and to compute the anomalous dimensions for higher spin currents of mixed symmetry. We also would like to study B-type Vasiliev theory dual to the theory of free fermions as in \cite{Sezgin:2003pt}.
For the application to the ABJ triality as mentioned in the introduction, it is necessary to examine marginal deformations by coupling Chern-Simons gauge fields to the free bosons or fermions as in \cite{Aharony:2011jz,Giombi:2011kc}, see also \cite{Maldacena:2012sf}.

\subsection*{Acknowledgements}

We are grateful to T.~Creutzig and P.~ B.~R{\o}nne for previous collaborations.
This work was supported in part by JSPS KAKENHI Grant Number 24740170.

\appendix

\section{Formulas}
\label{AppA}

In this appendix we summarize formulas used in the main context.

\subsection{Integrals}

\label{Kazakov}

During the computation we frequently move to the momentum basis \eqref{momentum}. 
For this purpose we use the integral
\begin{align}
\int d^3 x \frac{e^{i k \cdot x}}{|x|^{2 \Delta}} 
 = 2^{3 - 2 \Delta} \pi^{3/2} a(\Delta)  |k|^{2 \Delta -3} \, , \quad a(\Delta) = \frac{\Gamma (\frac32 - \Delta)}{\Gamma (\Delta)} \, ,
 \label{Fourier}
\end{align}
or equivalently 
\begin{align}
\frac{1}{|x|^{2 \Delta}} = 2^{-2 \Delta} \pi^{- 3/2} a(\Delta) \int d ^3 k \frac{e^{i k \cdot x}}{|k|^{3 - 2 \Delta}} \, .
\label{formula}
\end{align}
We also use the expressions with replacing $x$ and $k$.

Using the momentum basis, we can  show the following rules for calculating Feynman diagrams 
(see, e.g., \cite{Kazakov:1983ns}).
The first one is 
\begin{align}
\int d ^3 x_3 \frac{1}{|x_{13}|^{2 \alpha_1}|x_{23}|^{2 \alpha_2}}  = v(\alpha_1 , \alpha_2 , \alpha_3 )  \frac{1}{|x_{12}|^{2 \alpha_1 + 2 \alpha_2 - 3}}\, ,
\label{line}
\end{align}
where 
\begin{align}
 \quad v(\alpha_1 , \alpha_2 , \alpha_3) = \pi^{3/2} \prod_{i=1}^3 a (\alpha_i) \, , \quad 
 \alpha_3 =  3 - \alpha_1 - \alpha_2 \,  .
\end{align}
The second one with $\alpha_1 + \alpha_2 +\alpha_3 = 3$ is 
\begin{align}
 \int d ^ 3 x_4 \frac{1}{|x_{14}|^{2 \alpha_1}|x_{24}|^{2 \alpha_2}|x_{34}|^{2 \alpha_3}} 
 = v(\alpha_1 , \alpha_2 , \alpha_3)
  \frac{1}{|x_{23}|^{3 - 2 \alpha_1}|x_{31}|^{3  - 2\alpha_2}|x_{12}|^{3 - 2 \alpha_3}} \, .
  \label{vertex}
\end{align}

In order to read off the anomalous dimensions, we extract the contributions  proportional to $\log |x|$, thus we need the formulas involving the terms with $\log |x|$.
Taking derivative of \eqref{Fourier} with respect to $\Delta$, we find
\begin{align}
\int d^3 x \frac{e^{i k \cdot x}}{|x|^{2 \Delta}} \log |x|
 = - 2^{2 - 2 \Delta} \pi^{\frac32} a(\Delta)  |k|^{2 \Delta -3} \left[ 
  - 2 \log 2 + \psi (\tfrac32 - \Delta ) - \psi (\Delta) + 2 \log |k| \right]
  \, .
  \label{Flog}
\end{align}
Setting $\Delta \to 0$, we have
\begin{align}
\int d^3 x \, e^{i k \cdot x}\log |x|
 = - 2 \pi^{2}  |k|^{ -3} 
 \label{Flog0}
  \, ,
\end{align}
where we have used 
\begin{align}
\left. \frac{d}{d \Delta} \left( \frac{1}{\Gamma (\Delta)}\right)  \right|_{\Delta \to 0}
 = \left. \frac{d}{d \Delta} \left( \frac{\sin \pi \Delta}{\pi} \Gamma (1 - \Delta) \right)  \right|_{\Delta \to 0} = 1 \, .
 \end{align}
Furthermore we obtain
\begin{align}
\int d^3 x \, e^{i k \cdot x}(\log |x|)^2
 = 2 \pi^{2}   |k|^{ -3} \left[ 
  - 2 \log 2 + \psi (\tfrac32  )  + 2 \log |k| - \psi (1) \right] 
  \label{Flog2}
\end{align}  
by taking the  second derivative and setting $\Delta \to 0$. 
Notice that
\begin{align}
 \left. \frac{d^2}{d \Delta^2 } \left( \frac{1}{\Gamma (\Delta)}\right)  \right|_{\Delta \to 0}
  = \left. \frac{d^2}{d \Delta^2} \left( \frac{\sin \pi \Delta}{\pi} \Gamma (1 - \Delta) \right)  \right|_{\Delta \to 0} = - 2 \psi (1) \, . 
\end{align}

Taking the derivative of \eqref{vertex} with respect to $\alpha_1$ and setting $\alpha_1 = 0$, we find
\begin{align}
\label{vertexlog}
\int d ^3 x_4 \log |x_{14}| \frac{1}{|x_{24}|^{2 \alpha_2}|x_{34}|^{2 \alpha_3}}
 =  - \frac{\pi^2}{4} \frac{\Gamma (\frac 32 - \alpha_2) \Gamma (\frac32 - \alpha_3)}{\Gamma (\alpha_2) \Gamma (\alpha_3)}
  \frac{1}{|x_{23}|^3 |x_{31}|^{3 - 2 \alpha_2} |x_{12}|^{3 - 2 \alpha_2}} \, .
\end{align}
Here we have used 
\begin{align}
\frac{d}{d \alpha_1} v(\alpha_1 ,\alpha_2 , \alpha_3) |_{\alpha_1 = 0}
 &= \pi^{3/2} \frac{\Gamma (\frac 32 )  \Gamma (\frac 32 - \alpha_2) \Gamma (\frac32 - \alpha_3)}{\Gamma (\alpha_2) \Gamma (\alpha_3)} \frac{d}{d \alpha_1} 
 \left. \left( \frac{1}{\Gamma (\alpha_1) } \right)\right|_{\alpha_1 = 0}  \\
 & = \frac{\pi^2}{2} \frac{\Gamma (\frac 32 - \alpha_2) \Gamma (\frac32 - \alpha_3)}{\Gamma (\alpha_2) \Gamma (\alpha_3)} \, . \nonumber
\end{align}

\subsection{Series}
\label{formulas}

We use the following sum formulas, which are  checked by Mathematica at least for small $s$.%
\footnote{We confirmed \eqref{formula1}, \eqref{formula10} for all $s$ and \eqref{formula8} for $s=2,4,\ldots,50$. We checked the other formulas for $s=2,4,\ldots, 150$.}
We need
\begin{align}
\sum_{n=0}^{\text{max}(l,s-l)} \frac{2^{2n}}{(2n)! (l-n)! (s-n-l)!}  = \frac{(2s)!}{s! (2l)! (2s - 2l)!} 
\label{formula1}
\end{align}
and
\begin{align}
\sum_{m,n = 0}^s \frac{(-1)^{m+n} \Gamma (m+n + \frac12) \Gamma (2s - m - n + \frac12)}{(2m)! (2s-2m)! (2n)! (2s - 2n)!}
  = c_s\frac{\pi}{2 (2s)!}\, .
  \label{formula15}
\end{align}
Here $c_0 = 2$ and $c_s =1$ for $s = 2,4, \ldots$. 
We also use
\begin{align}
 \label{formula7}
&\sum _{m,n=0}^s \frac{(-1)^{m+n}}{(2 m)! (2 n)! (2 s-2 m)! (2 s-2 n)!}\sum _{k=0}^m \sum _{l=0}^{s-n} \Gamma (k+1) \Gamma (l+1) \binom{m}{k} \binom{s- n}{l}
  \\& \times \frac{ \Gamma \left(-m+s+\frac{1}{2}\right) \Gamma \left(n+\frac{1}{2}\right)  \Gamma \left(s-k-l+m-n-\frac{1}{2}\right) \Gamma \left(s + k+l-m+n+\frac{3}{2}\right)}{\Gamma \left(k-m+s+\frac{3}{2}\right) \Gamma \left(l + n +\frac{3}{2}\right)} \nonumber \\
  &= \frac{2 \pi}{(2s-1) (2s+1)!} \, .
\nonumber
 \end{align}

Let us define a complicated function by
 \begin{align}
 \label{Hf}
  & H (s;n,m;k,l;p,q) = \frac{(-1)^{m+n}} {(2 m)! (2 n)! (2 s-2 m)! (2 s-2 n)!}
   \binom{s-m}{k} \binom{s-n}{l}
   \\ & \times 
  \frac{ \Gamma \left(m+\frac{1}{2}\right) \Gamma \left(n+\frac{1}{2}\right) \Gamma \left(k-\frac{1}{2}\right)  \Gamma \left(l-\frac{1}{2}\right) \Gamma \left(p+\frac{1}{2}\right) \Gamma \left(q+\frac{3}{2}\right) \Gamma (-l-n+s+1)  }{\Gamma (l+n) \Gamma \left(k+m+q+\frac{3}{2}\right) \Gamma \left(-l-n+p+s+\frac{3}{2}\right)} 
    \binom{l+n}{p} 
\nonumber   \\ & \times
    \binom{s-k-m}{q} \Gamma \left(l-k -m+n-p-q+s-\frac{1}{2}\right) \Gamma \left(k-l+m-n+p+q+s+\frac{3}{2}\right) \, .
    \nonumber 
 \end{align}
Then we can show
 \begin{align}
\sum _{m,n=0}^s \sum _{k=0}^{s-m} \sum _{l=0}^{s-n} \sum _{p=0}^{l+n} \sum _{q=0}^{s -k-m} H (s;n,m;k,l;p,q) = \frac{4 s \pi^3}{ (2s - 1) (2s + 1)!} \, .
 \label{formula8}
 \end{align}
 Moreover, we find
  \begin{align}
&   \sum _{m=0}^s \sum _{k=0}^{s-m} \sum _{q=0}^{s-k-m} H (s;0,m;k,0;0,q) = 0   \, , \quad
  \sum _{n=0}^s   \sum _{l=0}^{s-n} \sum _{p=0}^{l+n} \sum _{q=0}^{s} H (s;n,0;0,l;p,q) = 0  \label{formula9}
\end{align}
and
\begin{align}
\sum _{q=0}^{s} H (s;0,0;0,0;0,q) = 0   \label{formula10}\, .
  \end{align}

\section{From bulk Witten diagram to conformal perturbation theory}
\label{AppB}

In the main context, we have computed the anomalous dimensions in the conformal perturbation theory and identified each contribution to bulk Witten diagram. 
In this appendix, we go in the opposite direction. Namely, we start from the computation with bulk Witten diagrams and map to that in the boundary conformal perturbation theory.
A point was to rewrite the bulk scalar propagator with Dirichlet boundary condition in terms of that with Neumann boundary condition and extra boundary operator insertions as claimed in \cite{Witten:2001ua}. We begin with deriving this fact by explicitly rewriting the scalar propagator by slightly modifying the argument in \cite{Creutzig:2015hta}.

Let us denote the bulk and boundary coordinates as $X_i$ and $x_i$, respectively.
We also represent the bulk-to-bulk, bulk-to-boundary, and boundary-to-boundary propagators for the scalar field with dual dimension $\Delta$ as
$\Pi_{\Delta} (X_1 , X_2)$, $\Pi_{\Delta} (X , x)$, and $\Pi_{\Delta} (x_1,x_2)$.
We use the normalization of the bulk scalar so that the kinetic term is of the standard form.
If we assume the coupling to boundary scalar operator $\tilde{\mathcal{O}}_\Delta$ as
$\int d^d x \phi \tilde{\mathcal{O}}_\Delta$, then the present normalization implies that
\begin{align}
\langle \tilde{\mathcal{O}}_\Delta (x_1) \tilde{\mathcal{O}}_\Delta (x_2) \rangle
= \Pi_{\Delta} (x_1 , x_2) = \frac{N_\Delta}{|x_{12}|^{2 \Delta}} \, , \quad N_\Delta =
\frac{(2 \Delta - d) \Gamma (\Delta)}{\pi^{d/2}\Gamma (\Delta - d/2)} \, . 
\end{align}
For $d=3$  and $\Delta = 1,2$, we have $N_1 = 1/(2 \pi^2)$ and $N_2 = 1/\pi^2 $.
Comparing with \eqref{scalar2pt}, we should rescale as
\begin{align}
\tilde{\mathcal{O}}_1 = \frac{1}{2 \pi N^{1/2}} \mathcal{O} \, .
\end{align}
We also change the perturbation parameter as 
$\tilde f = 4 \pi^2 N f$ such that the deformation \eqref{def} becomes
\begin{align}
 \Delta S = \frac{f}{2} \int d^3 x \, \mathcal{O} (x) \mathcal{O} (x) =
 \frac{\tilde f}{2} \int d^3 x \, \tilde{\mathcal{O}}_1 (x) \tilde{\mathcal{O}}_1 (x)\, .
 \label{tdef}
\end{align}
With this notation we have from \eqref{sum1}
\begin{align}
\langle \tilde{\mathcal{O}}_1 (x_1)  \tilde{\mathcal{O}}_1 (x_2) \rangle_{\tilde f}
 \sim \frac{1}{\tilde f}  \delta^{(3)} (x_{12}) + \frac{1}{\tilde f^2} 
 \langle \tilde{\mathcal{O}}_2 (x_1)  \tilde{\mathcal{O}}_2 (x_2) \rangle_{0} 
 \label{tsum1}
\end{align}
for $\tilde f \to \infty$.

Next we compute the scalar propagator with Neumann boundary condition dressed by the boundary operator insertions \eqref{tdef}. The propagator can be written as
\begin{align}
 \Pi_{1}(X_1 , X_2) _{\tilde f} \equiv &
 \Pi_{1}(X_1 , X_2) 
 - \tilde f \int d^3 x  \Pi_{1}(X_1 , x)  \Pi_{1}(x , X_2) \\
 &+ \tilde f^2 \int d^3 x_1  d^3 x_2  \Pi_{1}(X_1 , x_1) \Pi_{1} (x_1 , x_2) \Pi_{1}(x_2 , X_2) + \cdots \, . \nonumber 
\end{align}
Summing over higher order corrections, we have
\begin{align}
 \Pi_{1}(X_1 , X_2) _{\tilde f} = &
 \Pi_{1}(X_1 , X_2) 
 - \tilde f \int d^3 x  \Pi_{1}(X_1 , x)  \Pi_{1}(x , X_2) \\
 &+ \tilde f^2 \int d^3 x_1  d^3 x_2  \Pi_{1}(X_1 , x_1)  \langle \tilde{\mathcal{O}}_1 (x_1)  \tilde{\mathcal{O}}_1 (x_2) \rangle_{\tilde f} \Pi_{1}(x_2 , X_2) \, .\nonumber 
\end{align}
Using \eqref{tsum1}, we have 
\begin{align}
\label{propmiddle}
 \Pi_{1}(X_1 , X_2) _{\tilde f \to \infty} = &
 \Pi_{1}(X_1 , X_2) + \Pi_\text{mix} (X_1 , X_2) 
\end{align}
with
\begin{align}
\Pi_\text{mix} (X_1 , X_2) \equiv
  \int d^3 x_1  d^3 x_2  \Pi_{1}(X_1 , x_1)  \Pi_{2} (x_1 , x_2) \Pi_{1}(x_2 , X_2) \,
 .\nonumber 
\end{align}
The integral over the boundary coordinates $x_1,x_2$ can be preformed as (see (5.8) of \cite{Creutzig:2015hta})
\begin{align}
 \Pi_\text{mix} (X_1 , X_2) = \Pi_{2} (X_1 , X_2) - \Pi_{1} (X_1 , X_2) \, .
\end{align}
Plugging this expression into \eqref{propmiddle} we obtain
\begin{align}
 \Pi_{1,0}(X_1 , X_2) _{\tilde f \to \infty} = \Pi_{2,0} (X_1 , X_2)
\end{align}
as claimed. Note that this relation holds with the correct normalizations.

Let us apply this fact into the evaluation of the Witten diagram in fig.~\ref{2pt}.
For the loop there are two propagators along the upper and lower lines.
Since the Vasiliev theory includes gauge fields and a scalar field, there could be the following three cases; 
\begin{itemize}
\item[(1)] Gauge fields run along both lines.
\item[(2)] Gauge fields run along only one line and a scalar runs along the other line.
\item[(3)] Scalar runs along both lines.
\end{itemize}
Now we assign the Dirichlet boundary condition for the scalar field, and the propagator can be divided into two parts as $\Pi_{2} = \Pi_{1} + \Pi_\text{mix}$.
Therefore, the cases (2) and (3) can be divided further;
\begin{itemize}
\item[(2a)] The scalar propagator is $\Pi_{1}$.
\item[(2b)] The scalar propagator is $\Pi_\text{mix}$.
\item[(3a)] Both propagators are $\Pi_{1}$.
\item[(3b)] One of two propagators is  $\Pi_{1}$ and the other is $\Pi_\text{mix}$.
\item[(3c)] Both propagators are $\Pi_\text{mix}$. 
\end{itemize}
The sum of (2b) and (3b) precisely maps to the integral corresponding to the left diagram of fig.~\ref{critical2}. Moreover, (3c) maps to the integral corresponding to  the right diagram of 
fig.~\ref{critical2}. The sum of the other parts, (1), (2a), and (3a), corresponds to the Witten diagram for the case with Neumann boundary condition assigned to the scalar field. So we can use the fact that there is no anomalous dimension from the diagram.%
\footnote{In \cite{Chang:2012kt} higher spin symmetry is shown to be preserved for A-type Vasiliev theory with $\Delta = 1$ scalar boundary condition. The no-go theorem in \cite{Maldacena:2011jn} implies that the theory is dual to free boson theory, where the higher spin symmetry is exact. Even so,
it would be nice if we can confirm this by directly evaluating the bulk one-loop diagram.}
In this way, we have succeeded to reproduce the integrals in the boundary conformal perturbation theory from the bulk Witten diagram in fig.~\ref{2pt}.

The argument on the Goldstone modes in \cite{Girardello:2002pp} suggests that there is no contribution to the anomalous dimensions from (3) or the sum of (3b) and (3c). 
In order to show this, we need to separate the contribution (3b) from the sum of (2b) and (3b).
Since we know only the sum from tree Witten diagrams or dual boundary correlation functions,
we may need to evaluate the loop integral directly for the purpose. 
It is also suggested that there is no contribution for spin $s$ current from the bulk Witten diagram with spin $s' (> s)$ bulk propagator, but we are unable to show this with the current method.

%\bibliographystyle{JHEP}
%\bibliography{AdS3}

\providecommand{\href}[2]{#2}\begingroup\raggedright\endgroup

\end{document}